# Theories of Hypergraph-Graph (HG(2)) Data Structure

Shiladitya Munshi
Meghnad Saha Institute of Technology
Kolkata, India
Web Intelligence and
Distributed Computing research Lab
Golfgreen, Kolkata: 700095, India
Email:shiladitya.munshi@yahoo.com

Ayan Chakraborty
Techno India College of Technology
Kolkata, India
Web Intelligence and
Distributed Computing research Lab
Golfgreen, Kolkata: 700095, India
Email: achakraborty.tict@gmail.com

Debajyoti Mukhopadhyay
Maharastra Institute of Technology
Pune, India
Web Intelligence and
Distributed Computing research Lab
Golfgreen, Kolkata: 700095, India
Email: debajyoti.mukhopadhyay@gmail.com

*Abstract*—Current paper introduces a Hypergraph-Graph (HG(2)) model of data storage which can be represented as a hybrid data structure based on Hypergraph and Graph. The proposed data structure is claimed to realize complex combinatorial structures. The formal definition of the data structure is presented along with the proper justification from real world scenarios. The paper reports some elementary concepts of Hypergraph and presents theoretical aspects of the proposed data structure including the concepts of Path, Cycle etc. The detailed analysis of weighted HG(2) is presented along with discussions on Cost involved with HG(2) paths. *Keywords—ypergraph, Graph, Hyperpath, Hyperedges, HG(2), Cost of HG(2) Pathypergraph, Graph, Hyperpath, Hyperedges, HG(2), Cost of HG(2) PathH*

## I. INTRODUCTION

Hypergraphs, a generalization of Graph, is being widely and deeply investigated since last few decades as a successful tool to represent and model complex concepts and structures in various areas of Computer Science and Discrete Mathematics. Graphs on the other hand have enjoyed the spotlight of heterogeneous research as an important and implementable tool to represent and analyze real world complex scenarios.

Hence, both Hypergraphs and Graphs are the well suited candidates for representing complex structures having logical or real world relations with certain advantages over each other. While Hypergraphs have more potential than graphs in representing complex scenarios, it lacks readability, simplicity and easeness in conceptualization and physical representation.

In the current paper, a new data structure "Hypergraph - Graph" (HG(2)) has been introduced, which takes the advantages of both the data structures of Graph and Hypergraph. This data structure can model a problem space into two logical partitions with different levels of complexities. The more complex level can be represented by Hypergraph in order to take the leverages of its power to designate the relationships among a group of objects. The other level which is characterized by comparatively lesser complexity and better orderedness, could be represented by Graph in order to take the advantage of its naturalness and simplicity.

With a high objective of introducing HG(2), the present paper has been organized as follows -
Section 2 proposes HG(2) and related theoretical aspects formally with definitions and illustrative examples. Section 3 reports the concepts of paths within HG(2) with some co related issues, followed by discussions on the impact of weights on HG(2) and computation of Cost of any HG(2) Path in Section 4. Lastly, Section 5 concludes the current investigation emphasizing the level of contribution , importance and future research scope related with study of HG(2).

## II. HYPERGRAPH - GRAPH (HG(2)) DATA STRUCTURE

Hypergraph - Graph data structure denoted as HG(2) is conceptualized as a model to represent a complex problem space based on certain criteria. The criteria could be formalized as follows -
The problem space ($P\ S$) must logically be divided into two levels with different complexities, one ($P\ SG$) with relatively lesser complexities, better orderdness and bounded by for-malized set of rules, and another ($P\ SH$) which could be characterized by greater complexities and absence of ordered rule sets.
The some or all interrelationship between objects of $P\ SH$ must be dictated by the objects of $P\ SG$ and even the behavior of some or all objects of $P\ SH$ must be defined by the objects of $P\ SG$. Here the term "object" is being used informally and must not necessarily indicate any Object Oriented paradigm. As the complex real life combinatorial structures are not rare at all, proposed HG(2) has an intrinsic objective to represent $P\ SH$ with Hypergraphs and $P\ SG$ with Graphs. The inter dependencies of $P\ SH$ and $P\ SG$ form the basis of evolution of the behavior of HG(2) as a whole.

The theories behind the Hypergraph Data Structure are presented in [5] and due to space constraint, it is omitted in the current discussion. On this background, following subsection presents the theories of Hypergraph - Graph (HG(2)) data structure directly.

### A. Introducing HG(2)

A Hypergraph-graph data structure HG(2) is a triple denoted as $HG(2) = (H, G, C)$ where $H$ is a Hypergraph, $G$ is a graph and $C$ is a set of *connectors*.

$H$ is a Hypergraph defined as $H = (V^h, E^h)$, where $V^h = v_1^h, v_2^h, \cdots, v_n^h, n = 2, 3, \cdots$; and $E^h = E_1^h, E_2^h, \cdots, E_m^h,$

$m = 2, 3, \cdots$ where each $E_i^h \subseteq V^h$

$G$ is a Graph defined as $G = (V^g, E^g)$, where $V^g = v_1^g, v_2^g \cdots v_p^g$, $p = 2, 3, \cdots$; and $E^g = e_1^g, e_2^g \cdots e_q^g$, $q = 2, 3, \cdots$ where each $e_i^g$ could be expressed in the form of $e_{xy}^g$ which connects $v_y^g$ from $v_x^g$.

$C$ is a set of *connectors*, which could be conceptualized as a set of all the dependencies between $P\,SH$ and $P\,SG$ (as described earlier) which are characterized by $H$ and $G$ respectively.

here we define two types of connectors:
(a) $c_{xy}^v$ which connects a node in the Graph $v_y^g$ from a node in the Hypergraph $v_x^h$. It is to note that the behavioral dependency of an object of $P\,SH$ on an object of $P\,SG$ gets realized through $c_{xy}^v$; and
(b) $c_{xy}^e$ which connects a node in the Graph $v_y^g$ from a Hyperedge $E_x^h$. Here $c_{xy}^e$ realizes the dependencies of collective behavior (bound with a specific relation) of the objects of $P\,SH$ on the objects of $P\,SG$.

The set of all $c_{xy}^v$s is noted as $C^v$ while $C^e$ represents all $c_{xy}^e$s. Hence on the basis of ongoing discussion, it could be concluded that $C = (C^v, C^e)$. For the rest of the paper, it is assumes for simplicity that the dependency flows from $P\,SH$ to $P\,SG$. That is the Hypergraph layer is dependent on the Graph layer. No dependency flows through a connector backward from the Graph layer to the Hypergraph layer. Above discussion can be illustrated with the example as shown in Fig 1. In this figure, an HG(2) is shown to have a Hypergraph $H$ and a Graph $G$. The $H$ and the $G$ are connected with *connectors* $C$. The Hypergraph $H$ is composed of $(V^h, E^h)$ where $V^h$ consists nodes 1, 2, 3, 4, 5, 6 & 7 and $E^h$ consists $E_1$, $E_2$, $E_3$ & $E_4$. Further, the Graph $G$ is composed of nodes $a$, $b$, $c$, $d$, $e$ & $f$. For simplicity the edges of the $G$ are not highlighted.

It is to be noted that $E_1$ is composed of nodes 1, 2, 3, $E_2$ is composed of nodes 3, 4, 5, 6, $E_3$ is composed of nodes 4, 5, 7 and finally $E_4$ is composed of nodes 5, 6 and 7.

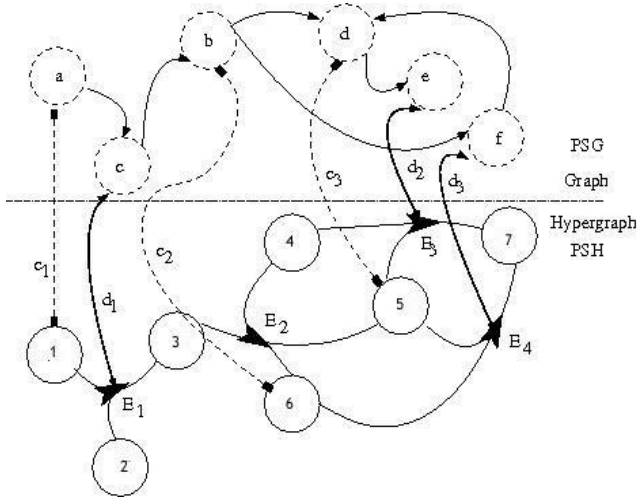

Fig. 1. Illustration of an HG(2) data structure

Following is the identification of head and tail nodes for each Hyperedge:
$H(E_1) = 1, 2$ and $T(E_1) = 3$
$H(E_2) = 3, 4$ and $T(E_2) = 5, 6$
$H(E_3) = 4, 5$ and $T(E_3) = 7$
$H(E_4) = 5, 6$ and $T(E_4) = 7$

The Fig 1 further illustrates that there are three connectors (marked with dashed square ended bi-directional arrow) that connect Hyperedge nodes and Graph nodes. They are $c_{1a}^v = c_1$, $c_{6b}^v = c_2$ and $c_{2d}^v = c_3$. Moreover there are three connectors (marked with bold bidirectional arrow) that connect a Hyperedge with a Graph node. They are identified as $c_{1c}^e = d_1$, $c_{3e}^e = d_2$ and $c_{4f}^e = d_3$. Hence formally $C$ which is a pair $(C^v, C^e)$ holds the following:
$C^v = c_1, c_2$ and $c_3$; and $C^e = d_1, d_2$ and $d_3$.

Though the diagram shows the individual connectors with bidirectional arrow, the earlier assumption still holds that the direction of the dependency is from Hypergraph to Graph layer. The bidirectional arrows have been given for easy diagrammatic identification of connectors out of many different edges.

However, Fig 1 shows the entire problem space to be logically divided (by a dashed horizontal divider) into $P\,SG$ and $P\,SH$ which are represented by the Graph and Hypergraph respectively.

On the basis of current discussion, next section reports critical issues related with concept of Path in a HG(2).

### III. CONCEPT OF PATH IN HG(2)

The study of Paths in HG(2) and the related algorithms demand formalization of certain essential terminologies which are presented next.

A *node dependent pair* or simply the *node pair* is defined as the pair of Hypergraph and graph nodes which share a dependency relation through $c_{xy}^v$ and is denoted as $v_x^h(v_y^g)$. In Fig 1, 1($a$), 6($b$) and 5($d$) could be identified as three *node pair*s. It is to be noted that, in context to a Hypergraph node not involved in $C^v$, the *node pair* is symbolized as $v_x^h(\,)$. The *node pair* of a Hypergraph node $n$ is designated as $NP^n$.

Similarly a *edge dependent pair* or simple *edge pair* is defined as the pair of Hyperedges and graph nodes which share a dependency relation through $c_{xy}^e$ and is denoted as $E_x^h(v_y^g)$. Any *edge pair* in context to a Hyperedge not involved in $C^e$ is symbolized as $E_x^h(\,)$. In Fig 1, three *edge pair*s could be identified as $E_1(c)$, $E_3(e)$ and $E_4(f)$. However, the *edge pair* of a Hyperedge $n$ is designated as $EP^n$.

Based on the terminologies discussed above, A path $P_{st}^{HG(2)}$ in HG(2) (known as *HG(2) path* of length $q$ could be defined as a sequence of node pair and edge pair. That is $P_{st}^{HG(2)} = \{NP_1^{V_{ih}=s}, EP_1^{E_{ih}}, NP_2^{V_{ih}}, EP_2^{E_{ih}} \cdots EP_q^{E_{ih}}, NP_{q+1}^{V_{ih}=t}\}$;
where following conditions are all true;
i) $P_{st}^{HG(2)}$ is a valid Hyperpath considering only the Hyperedges in $H$
ii) each $NP^{V_{ih}} = v_x^h(v_y^g)$ and each $EP^{E_{ih}} = E_z^h(v_w^g)$; i and x $= 2, 3, \cdots$

iii) for any $NP_i^{v_h}$, $EP_i^{E_h}$ pair, there exists a valid path from $v_Y^g$ to $v_W^g$ in $G$

iv) for any $EP_i^{E_h}$, $NP_i^{v_h}$ pair, there exists a valid path from $v_W^g$ to $v_Y^g$ in $G$

A path in Graph $G$ traced by the Hyperpath in Hypergraph $H$ is defined as *Graph Path* or simple as *GPath*. A *GPath* is formed by the graph nodes involved in constituent *node pairs* and *edge pairs* (in sequence) of a *HG(2) path*.

An *HG(2) path* $P_{st}^{HG(2)}$ is said to contain a *Graph Loop* or simply *GLoop* if there exists a cycle in *GPath*

An *HG(2) path* $P_{st}^{HG(2)}$ is said to contain a *Hypergraph Loop* or simply *HLoops* if there exists a cycle in Hyperpath considering the Hypergraph only.

An *HG(2) path* $P_{st}^{HG(2)}$ containing *GLoop* must not necessarily contain *HLoop* and the vice-versa is also true. Even the *Gloop* and *Hloop* may co-exist in a single $P_{st}^{HG(2)}$.

An *HG(2) path* $P_{st}^{HG(2)}$ is qualified as *Elementary* if all the nodes in the Hyperpath are distinct, and the same is termed as *Simple* if all the Hyperedges are distinct.

However, the idea regarding HG(2) path, that evolved up from the ongoing study, could be illustrated with the example shown in Fig 1.

Let us consider an *HG(2) path* $P_{17}^{HG(2)}$. This path starts from Hypergraph node 1 ($V_1^h = 1$) and reaches another Hypergraph node 7 ($V_7^h = G$). $P_{17}^{HG(2)}$ follows two Hyperpaths The nodes and edges of the first and second Hyperpaths result sequences of *node pair* and *edge pair* as shown below:

{1(*a*), $E_1$(*c*), 3( ), $E_2$( ), 5(*d*), $E_3$(*e*), 7( )} and;
{1(*a*), $E_1$(*c*), 3( ), $E_2$( ), 6(*b*), $E_4$(*f*), 7( )}

Both these *HG(2) paths* ensure that the every following graph node in the sequence are connected to its previous node. For example in {*a, c, d, e*}, *c* is connected to *a*, *d* is connected to *c* (through *b*) and *e* is connected to *d* That is {*a, c, b, d, e*} and {*a, c, b, f* } form two valid paths in Graph *G*. Hence these two graph paths can be termed as *GPath*.

There exists another Hyperpath {1, $E_1$, 3, $E_2$, 5, $E_4$, 7} that can be followed for $P_{17}^{HG(2)}$. The corresponding *node pair* and *edge pair* sequence is generated as {1(*a*), $E_1$(*c*), 3( ), $E_2$( ), 5(*d*), $E_4$(*f*), 7( )} and the sequence of Graph nodes corresponding the *node pair* and *edge pair* sequence is generated as {*a, c, d, f* }. It is to be noted here that node *f* is not connected to node *d* and hence this sequence can not form a valid path in the Graph *G*. Hence {1(*a*), $E_1$(*c*), 3( ), $E_2$( ), 5(*d*), $E_4$(*f*), 7( )} is not a valid *HG(2) path*.

A path in a HG(2), thus, as seen in the above discussion, a valid Hyperpath with some additional constraints. These additional constraints can be attributed to the *GPath*. Hence it is to be considered that an HG(2) path can have greater potential than a Hypergraph to model sequence of complex relationships.

On this background, the next section discusses the notions of Weighted HG(2) and consequently the cost involved in HG(2) paths.

IV. COST ANALYSIS OF PATH ON A WEIGHTED HG(2)

A Weighted HG(2) is composed of a weighted Hypergraph, a weighted Graph and weighted Connectors. In this inves-tigation, the weight simply means the weight of edges that means the cost of connecting nodes within individual Graph, Hypergraph, nodes of Graph an Hypergraph and nodes of Graph and edges of Hypergraph. The weight of graph edges that represents the cost of connecting two Graph nodes $v_i^g$ and $v_j^g$ is denoted by $C_{e_{ij}^g}$ or $C_{e_k^g}$ where $e_k^g$ connects node *i* and node *j*. Similarly, the weight of hyperedge that represents the cost of connecting $v_a^h$, $v_b^h$, $v_c^h$ · · · related by hyperedge $E_i^h$ is denoted by $C_{E_{ih}}$. The weight of node to node connector $c_{ij}^v$ is the cost of connecting a node of Hypergraph $v_i^h$ and a node of graph $v_j^g$, and it is denoted as $C_{c_{ij}^v}$. Hence $C_{c_{ij}^e}$ will determine the weight of edge to node connector $c_{ij}^e$ which connects $E_i^h$ to $c_j^v$. Based on this discussion, the cost associated with a path can be viewed as follows.

Before proceeding any further, it is to be noted that ant *HG(2) Path* may give rise to multiple valid Hyperpath routes. For example, *HG(2) Path* $P_{17}^{HG(2)}$ in the HG(2) shown in Fig 3 gives rise three valid Hyperpath routes {1, $E_1$, 3, $E_2$, 5, $E_3$, 7} , {1, $E_1$, 3, $E_2$, 6, $E_4$, 7} and {1, $E_1$, 3, $E_2$, 5, $E_4$, 7}. Consequently we get two valid HG(2) route out of these three Hyperpath routes as :
{1(*a*), $E_1$(*c*), 3( ), $E_2$( ), 5(*d*), $E_3$(*e*), 7( )} and;
{1(*a*), $E_1$(*c*), 3( ), $E_2$( ), 6(*b*), $E_4$(*f*), 7( )}

Each of these valid HG(2) routes will be denoted by $R^1_{P_{17}^{HG(2)}}$ and $R^2_{P_{17}^{HG(2)}}$. Hence in general, $R^k_{P_{st}^{HG(2)}}$ denotes the $k^{th}$ route for *HG(2) path* $P_{st}^{HG(2)}$.

It is to be further noted that for each individual valid HG(2) route, there may exist multiple *GPath*s. Each of these *GPath*s are the valid connectors among the subgraph of G which has been induced by corresponding $c_{ij}^v$ and $c_{ij}^e$ of the chosen HG(2) route $R^k_{P_{st}^{HG(2)}}$. Hence the notation $G_{R^k_{P_{st}^{HG(2)}}}$ will denote the subgraph of G induced by the valid HG(2) route $R^k_{P_{st}^{HG(2)}}$ of *HG(2) Path* $P_{st}^{HG(2)}$. An $i^{th}$ *GPath* $GP_{pq}$ on $G_{R^k_{P_{st}^{HG(2)}}}$ denotes a path from graph node *p* to graph node *q* where *p* and *q* are the first and last graph nodes encountered tracing a *HG(2) Path* $P_{st}^{HG(2)}$. The $i^{th}$ *GPath* for graph node pair *p* and *q* is thus denoted by $GP_{pq}^i$

The simplified concept of *HG(2) Path* presented in the previous section could be represented by single or multiple *HG(2) route*s where each such *HG(2) routes* can be traced with single or multiple *GPath*s. The cost of *HG(2) Path* hence will be dependent on cost of *HG(2) routes* and cost of *GPath*.

However Cost of *HG(2) route* is the summation of cost of Hyperedges that is $C_{R^k_{P_{st}^{HG(2)}}} = \sum_{i=0}^{n} C_{E_i^h}$ where $E_0^h$ to $E_n^h$ are the Hyperedges that constitute the Hyperpath of the *HG(2)* *route*. Similarly the cost of GPath $C_{GP_{pq^i}} = \sum_{i=0}^{n} C_{e_i^g}$ where $e_i^g$ to $e_n^g$ constitute the *GPath*.

This paragraph will investigate the effects of cost of connectors into the computation of cost of *HG(2) Path*. While considering $GP_{pq}^i$, there might be some nodes which are connected with Hypernodes and Hyperedges for the Hyperpath selected, and at the same time ther might exist some graph nodes $v_i^g$'s who are not connected with the any of the Hypernodes or Hyperedges, rather they got traced in order to establishing a valid path between two consecutive graph nodes of *node pair* and *edge pair* (discussed in previous section).

Let us consider the case as shown in Fig 3. For the *Hg(2) Path* $P_{st}^{HG(2)}$ and for *G(2)* route $R^T_{P_{st}^{HG(2)}}$, node *b* got selected to establish the path between node *c* and node *d*. Here node *c* and node *d* are directly connected to the Hyperedge $E_1$ and Hypernode 5 respectively, but node *b* has no connection with the selected Hyperpath. While the nodes connected with the selected Hyperpath are referred to as *Participating Nodes*, the other types of graph nodes are referred to as *Auxiliary Nodes*.

The characterization of *Auxiliary Nodes* results that there might be three cases as follows:
(i) A particular *Auxiliary Node* has no connection at all. Neither it participates in any $c^v_{ij}$ nor in any $c^e_{ij}$
(ii) A particular *Auxiliary Node* has connections out of the scope of the selected Hyperpath; and
(ii) A particular *Auxiliary Node* has connections within the scope of the selected Hyperpath.
The example shown just above represents the case (ii).

The engagement of Case (i) clearly has no effect in computing Cost of *HG(2) Path*, while engagement of Case (ii) and case (iii) induces some dependency of Graph layer to Hypergraph layer. But as stated earlier in Section 3, for simplicity, there exists no dependency from Graph layer to Hypergraph layer, so even the existence of Case (ii) and case (iii) does not make any difference in computation of Costs of *HG(2) Path*. Had there been a scope of mutual dependency between Graph and Hypergraph layer, the existence of Case (ii) and Case (iii) had have severe effect in *HG(2) Path* cost, but in present research, this is beyond the scope of any further discussions.

On this background, for all the *Participating Node*s, there are wighted connectors and only the connectors connected to the *Participating Node*s will contribute the cost in computing *HG(2) Path*. Hence the cost of connectors for a set of specific $R^x_{P_{st}^{HG(2)}}$ and specific $GP_{pq}^y$, which is denoted as $C_{xy}$, could be given by the relation:

$C_{xy} = \sum_{r=0}^{n} C_{C^{v_r}} + \sum_{r=0}^{n} C_{C^{e_r}}$ where each $c^v_r = c^v_{ij}$ and each $c^e_r = c^e_{ij}$

denote node to node and edge to node connector respectively which are connected to each of the participating node *j* for a specific selected $GP_{pq}^w$ ($w = 1, 2 \cdots$)

All the preceding discussions identify the cost expression of a *HG(2) Path* $P_{st}^{HG(2)}$ for a specific *HG(2) route* $R^u_{P_{st}^{HG(2)}}$ and specific *GPath* $GP^z$ as

$C_{P_{st}^{HG(2)}}(u, z) = C_{R^u_{P_{st}^{HG(2)}}} + C_{GP_{pq}^z} + C_{uz}$.

For a given $P_{st}^{HG(2)}$ there will be multiple set of (*u, z*) pair and the $min(C_{P^{HG(2)}_{st}}(u, z))$ will denote the least cost.

## V. CONCLUSION

The present research has introduced a data structure which has been denoted as HG(2). This proposed data structure has a great potential in modeling and resolving complex combinatorial problems. The basic Hypergraph and Graph based architecture of this data structure has been presented. The theoretical concepts of *Paths* in HG(2) has been discussed. Cost Analysis of *HG(2) Path* has been thoroughly presented.

Within the current discussion, the future scope of HG(2) research has aptly been identified to focus on generation of Minimum Spanning Tree out of HG(2), Identification of Least Cost Path, HG(2) Path Traversal algorithms and modeling of different real life problems (like RDF integration in Semantic Web).